\documentclass[cha,reprint,groupedaddress,twocolumn]{revtex4-1}
\usepackage[utf8]{inputenc}
\usepackage[T1]{fontenc}
\usepackage{amsmath}
\usepackage{amsfonts}
\usepackage{amssymb}
\usepackage{graphicx}
\begin{document}

\title{Low-cost experiment to measure the speed of light}

\author{Faraz Mehdi}
\address{Department of Physics, University of Mumbai, Santa Cruz (E),\\
Mumbai 400 098, India\\Department of Physics, Ramniranjan Jhunjhunwala College, \\ Ghatkopar (W), Mumbai 400 086, India\\farazmehdi786@gmail.com}

\author{Kiran M. Kolwankar}
\address{Department of Physics, Ramniranjan Jhunjhunwala College, \\ Ghatkopar (W), Mumbai 400 086, India\\Kiran.Kolwankar@gmail.com}

\date{\today}

%\onecolumn
%\begin{center}
%	\textbf{ {\Large ABSTRACT}}
%\end{center}

\begin{abstract}

In this paper, we demonstrate a low-cost method to measure the speed of light. It uses instruments which are readily available in any  undergraduate laboratory in a developing country and some components which are inexpensive. The method is direct as it measures the time of flight of the LASER beam and easy to implement. It will allow students to verify the finite value of the speed of light first hand. It can be part of the undergraduate syllabus as a regular experiment or a demonstration experiment.\newline
%The speed of light can be measured by using different ways, principles and apparatus, most of which are either too complicated to be performed in a lab in the time limit provided for a practical or they are too expensive to be brought as a regular experiment. In this paper we are proposing a method which is efficient not only in time but also in cost. It consists of mainly those components which either are already present in an undergraduate lab (in India) or are very inexpensive and readily available in the market.\newline
\end{abstract}


\maketitle

\section{Introduction}
Light travels exceedingly fast. This lead to the belief among the earlier researchers that it travels with infinite speed. To fathom the fact that it travels with finite speed, though very large as compared to the everyday objects, was difficult at those times and, even now, it poses a barrier among the young students studying science. As a  result, it would be of interest to have an experiment in which the students can themselves measure the speed of light directly and verify that it is finite.

There are different ways to estimate the speed of light but most of them are either hard to perform in the given time, e.g., the experiments consisting of a rotating mirror based on the principle proposed by Fizeau and Foucault, or they are too expensive e.g., the use of pulsed LASER to measure the speed of light. There are commercially available set-ups for undergraduate laboratories but they cost usually in several lakhs of rupees. We present here an experiment which measures the time of flight of a modulated laser beam over a distance. Given the instruments like an oscilloscope (or a DSO), power supplies and a signal generator, which are readily available in any undergraduate laboratory, the other components cost less than 500 INR.

% the speed of light experiment by Pasco[3] or Light velocity measuring instrument by Leybold[4].

%Being a science student, we have studied that the speed of light is finite and its value is around 3x10$^{8}$m/sec but imagining the same is a rather difficult task for a fresh undergrad student and the idea of finding this constant of nature inside a lab is not only fascinating but it also creates interest in them to pursue physics.\newline
% \newline 
 
There are also indirect ways to measure the speed of light. One example consists of detecting nodes created by the standing waves formed by the microwaves in a microwave oven in the form of blackened spots on, say, a piece of paper and then estimating the speed of the radiation using known frequency. This is not a direct way in the sense that it involves another concept and, moreover, the microwaves are not visible leading to lesser impact.
 %There is an option using the microwave to find the wavelength and then multiplying it by the frequency provided by the manufacturer to get the almost exact speed, but this approach is more theoretical than practical as it requires the knowledge of wave propagation , though it can be used as a demonstration experiment explaining the wave mechanics but is not a very good candidate for a regular experiment for speed of light estimation.\newline
 
% The method that we are proposing is based on the time of flight principle as in this method we divide light into two components using a beam splitter and then one of the beam will travel more distance than the other and the time-delay between the two is recorded. Then the ratio of excess distance to the time lag gives the speed of light.This method is both inexpensive and easy to be performed inside a lab. It incorporates some of the basic concepts from electronics and optics as well and the process of setting it up improves the lab skills of the student.The experimental setup is so simple that the students would know each and every part of it, with their working and importance, as no complicated component is required except the very basic lab equipment and the calculation is based on both the statistical mean and slope of the graph. Including the LASER, cost of the entire experiment is under 500 INR, so it can be arranged for all the students. It only requires a long lab around 10-15 m, which is not a problem in India.\linebreak

The need for a pedagogical experiment for direct estimation of the speed of light was realised more than 50 years ago~\cite{CMMW,Tyl,RMPA,VB}. As the technology evolved over time, these experiments too~\cite{EW,BHSS,DHH,JOS,Bea,AM,GRC,Peg,OM}. Most of these experiments involved a light source with modulated intensity which is allowed to travel some distance and the speed of light is estimated from the phase shift between the waveform detected at the source and at the end of the beam. A lamp, which was used in the initial set-ups, was replaced by LED, then a pulsed laser or a diode laser. Our aim was to design the experiment using instruments easily available in a typical undergraduate laboratory in a developing country and all the set-ups available in literature fail in this criterion for one or more reasons. For example, we can not use LED as the source of light as, in that case, distance travelled is small and then one needs to use very high frequency (> 50MHz). Such signal generators are not common. But the diode lasers are not at all expensive these days using which one can allow the beam to travel a longer distance thereby reducing the frequency to around 300 kHz. The set-up described in~\cite{OM} turns out to be very close to ours but there the authors have used an advanced oscilloscope and some video analysis using computers which is not the case in our experiment.

In the next section we describe our experimental set-up in detail. Then in the section~\ref{se:meas}, we discuss our measurements and in~\ref{se:res} we discuss the results. Finally, we end by some concluding remarks in~\ref{se:concl}
 
\section{Experimental set-up}\label{se:expt}

\begin{figure}
	\includegraphics[width=1\linewidth]{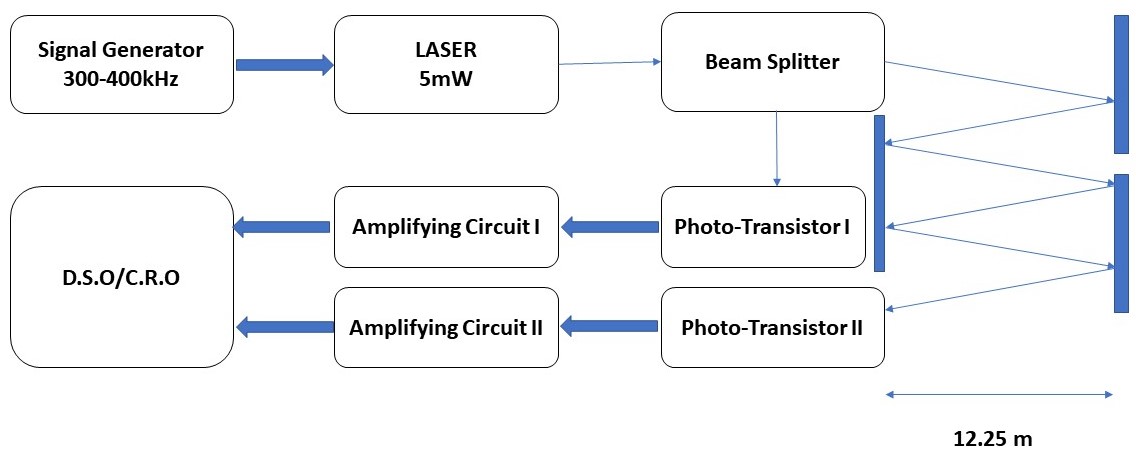}
	\caption{The block diagram}
	\label{fig:block}
\end{figure}

The block diagram of the experimental set-up is shown in the figure~\ref{fig:block}. We are using a red 5mw diode LASER module as source of light. It is powered by a square-wave generator at a frequency of around 300-400 kHz. This particular frequency range is chosen because we already know the value speed of light and we also know the distance we are allowing the light to travel. Hence we expect the time delay in hundreds of nanoseconds and to get a better measurement of the time-delay in this range on oscilloscope with more precision, the frequency should be in the range taken. This range can be changed depending upon the distance light is allowed to travel. \newline

\begin{figure}
	\includegraphics[width=1\linewidth]{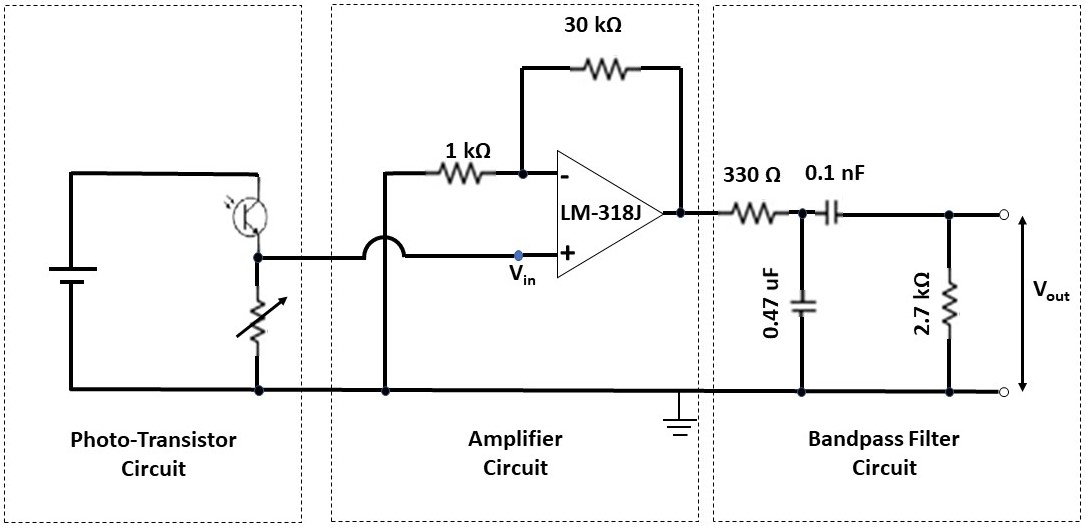}
	\caption{The cicuit diagram}
	\label{fig:circuit}
\end{figure}

The modulated light from the LASER falls on a thin glass slab which splits the beam into a reflected and a refracted part. The reflected light falls on a photo-transistor which is then connected in series with a resistance (as shown in the fig~\ref{fig:circuit}) and the voltage across the resistance gives the required signal. This signal from the receiver is weak and has noise. So to remove the noise and amplify the signal the two ends of the resistor are connected to the amplification  and noise cancellation circuit (see fig~\ref{fig:circuit}). For the amplification of the signal we use a  high-speed Op-Amp (LM318J) in non-inverting amplifier mode with a gain of 30. Since we are working with high frequencies (of the order of 300kHz) the Op-Amp also generates noise of high frequencies (in MHz). To remove this high frequency noise along with the 50Hz background noise coming from the mains, we connected the output of the Op-Amp to a band-pass filter allowing only the frequencies from 1kHz to 500kHz to pass and thus improving the signal to noise ratio. This signal is then observed on C.R.O or D.S.O for the measurement.\newline

The refracted part of light falls on a mirror (placed at sufficiently large distance so that the time delay of the signals can be observed on the C.R.O./D.S.O.) and gets reflected back near the experimental set-up to fall on a convex lens which converges it on the second photo-transistor with identical amplification and noise-cancellation circuit as that of the first photo-transistor. Then the output is observed on the second channel of the C.R.O. or D.S.O. To increase the distance travelled, it is possible to have several reflections from the mirrors at the farther end. This arrangement of multiple reflections allows us to take the readings for several distances.\newline

\section{MEASUREMENT}\label{se:meas}

We observe the signals from both the photo-transistors corresponding to reflected and refracted beam. The refracted beam travels more distance than the reflected one and hence will take more time which results in the shifting of its waveform towards right on the time axis and a difference between the peaks of the two signals can be observed. The speed of light is then measured by taking ratio of the excess distance travelled by the refracting light and this time difference between the peak. This delay can be observed even on a simple CRO which immediately confirms the finiteness of the speed of light. It also yields a value of the speed of light which is quite close to the actual value. For more accurate results, we use a DSO and take readings for several distances as described below.

Our first measurement was for zero relative distance as for 0 distance ideally there should be no time difference and any observed time difference would correspond to time-lag in the two receiver circuits. We observed that both the waveforms coincided so there was no observable time-lag in the two receiver circuits. Then different measurements were made by changing the position of the mirror and also by using more mirrors leading to multiple reflections. %Obviously the quality of mirror is important as the LASER signal can become broader with each reflection that is why we used a convex lens to focus the LASER point. %To minimize the error due to the additional distance added by the refraction through the lens we used a thin lens with focal length of around 30cm.

\begin{figure}
	\includegraphics[width=0.8\linewidth]{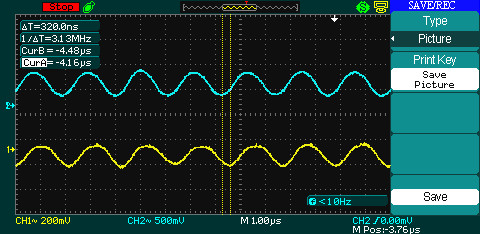}
	\includegraphics[width=0.8\linewidth]{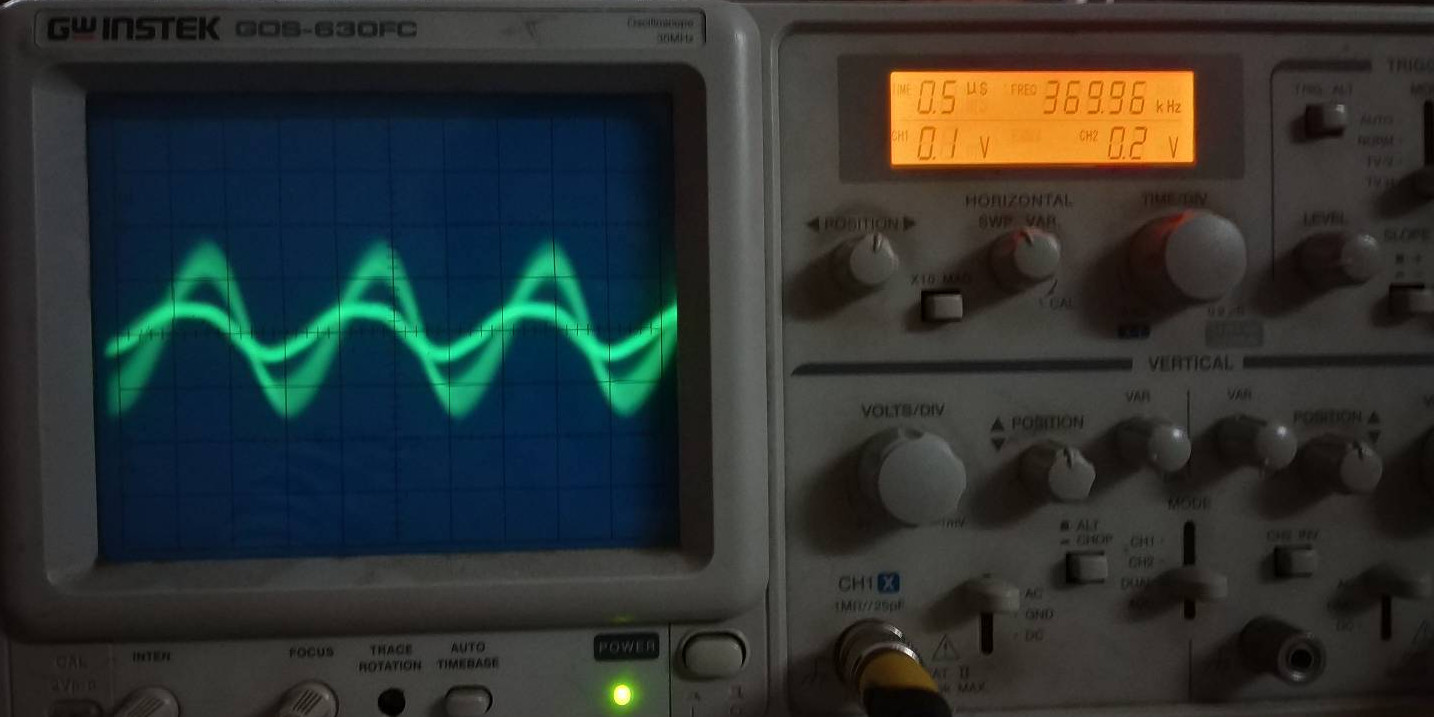}
	\caption{Image of the DSO and oscilloscope screen demonstrating the delay in two waveforms}
	\label{fig:wave}
\end{figure}

An example of the waveform that we see is depicted in Fig.~\ref{fig:wave}
We see that, though we have given a square wave power supply to the laser, the waveform is more near to a sinusoidal waveform. This happens because the higher frequencies get filtered out at the laser and/or at the photo-transistors. This is a result of their finite switching time.

\section{Results}	\label{se:res}
\begin{figure}
	\includegraphics[width=0.8\linewidth]{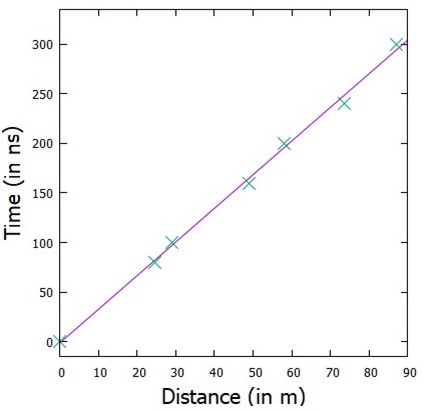}
	\caption{The graph of delay time vs the distance travel}
	\label{Graph}
\end{figure}

We carried out the experiment twice with three reflecting mirrors at the farther end. The distance of the farther mirror from the laser was different in each case. As a result, we generated values of delay times for 6 different distances. They are listed in a table below. The mean value of the speed of light turns out to be 2.98x$10^8m/s$. These are also plotted in the Fig~\ref{Graph}. We obtain a good straight line with slope 3.39x$10^{-9}s/m$. Taking the reciprocal of this slope gives us the speed of light to be 2.95x$10^8m/s$.

\begin{center}
	The following is our observation table:

\begin{tabular}{|c|c|}
	\hline 
	\rule[-1ex]{0pt}{2.5ex}\textbf{ Distance (in m)} & \textbf{Time Difference (in ns)}   \\
	\hline
	\rule[-1ex]{0pt}{2.5ex}  0& 0 \\
	\hline
	\rule[-1ex]{0pt}{2.5ex} 24.5 & 80  \\
	\hline
	\rule[-1ex]{0pt}{2.5ex} 29 & 100 \\
	\hline
	\rule[-1ex]{0pt}{2.5ex} 49 & 160  \\
	\hline
	\rule[-1ex]{0pt}{2.5ex} 58 & 200 \\
	\hline
	\rule[-1ex]{0pt}{2.5ex} 73.5 & 240  \\
	\hline
	\rule[-1ex]{0pt}{2.5ex} 87 & 300  \\
	\hline
\end{tabular}
\end{center}

\section{Concluding remarks}\label{se:concl}
We have demonstrated a simple and direct method for measurement of the speed of light in an undergraduate laboratory.
The method that we are proposing is based on the time of flight principle. In this method we divide light into two components using a beam splitter and then one of the beam is made to travel some large distance. The resulting time-delay between the two is recorded. Then the ratio of excess distance to the time lag gives the speed of light. This method is both inexpensive and easy to perform in an undergraduate laboratory. It incorporates some of the basic concepts from electronics and optics as well. The process of setting it up improves the laboratory skills of the student. The experimental set-up and the circuit is so simple that the students can build it from scratch. Thus they would know each and every part of it, with the working and importance. There is no special equipment required. The cost of components required is less than 500 INR which are reusable in other experiments. It is our experience that the students do experience a thrill when they see the finiteness of the speed of light for themselves for the first time.

\section*{Acknowledgements}

We would like to acknowledge the financial support from Department of Biotechnology, India under their Star College Scheme. A major part of this work was carried out as FM's undergraduate project in RJ College. Thanks are also due to several earlier students since 2011 who have rendered their help in the initial stages of this experiment.

\end{document}